\documentclass[11pt,a4paper]{article} 
\usepackage{amsfonts}
\usepackage[dvips]{graphicx}
\usepackage{amsthm}
\usepackage{amsmath}
\usepackage{epsfig}
\usepackage{t1enc}
\usepackage{amssymb}
\usepackage{latexsym}
\usepackage{bm}
\usepackage[hang]{caption}

\def\r{\rho} \def\d{\delta} \def\D{\Delta} 
  \def\H{{\cal H}}
\def\E{{\cal E}} \def\B{{\cal B}} 
\def\a{\alpha} \def\b{\beta} \def\tw{\tilde w}
\def\hc{\hat c} \def\0{{(0)}}  \def\z{\zeta}
\def\Bl{\Big(} \def\Br{\Big)} \def\BL{\Big[} \def\BR{\Big]}
\def\C{{\cal C}} \def\Pz{\Phi_0} \def\e{\epsilon}
\def\tim{\mbox{\scriptsize $\ \times\ $}}
\def\te{\tilde \eta} \def\tte{\tilde{\te}}
\def\tz{\tilde z} \def\ttz{\tilde{\tz}} \def\I{{II}}
\def\tx{\tilde x} \def\ttx{\tilde{\tx}}
\def\ty{\tilde y} \def\tty{\tilde{\ty}}
\def\tX{\tilde X} \def\tY{\tilde Y}
\def\hM{{\hat M}} \def\hN{{\hat N}}
\def\tm{\tilde \mu} \def\dP{\D \Phi} \def\Po{\Phi_{1,nd}}

\title{\bf Suppression of large-scale perturbations \\ by stiff solid}

\author{Vladim\'ir Balek\footnote{e-mail
address: balek@fmph.uniba.sk}\ \ and Matej \v
Skovran\footnote{e-mail address: skovran@fmph.uniba.sk}
\\
{\it Department of Theoretical Physics, Comenius University,
Bratislava, Slovakia}}

\begin{document}

\renewcommand{\figurename}{Fig.}
%\captionsetup[figure]{labelsep=newline}

\maketitle
\maketitle\abstract

{Evolution of large-scale scalar perturbations in the presence of
stiff solid (solid with pressure to energy density ratio $> 1/3$)
is studied. If the solid dominated the dynamics of the universe
long enough, the perturbations could end up suppressed by as much
as several orders of magnitude. To avoid too steep large-angle
power spectrum of CMB, radiation must have prevailed over the
solid long enough before recombination.}

%KAPITOLA 1

\section{Introduction}

In standard cosmology, large-scale perturbations stay unchanged
throughout the Friedmann expansion that started after inflation,
except for the last period before recombination when the Newtonian
potential was suppressed, due to the transition from radiation to
matter, by the factor 9/10 (see, for example, \cite{mukh}). The
potential is not affected even by phase transitions and
annihilations taking place in the hot universe, as long as the
matter filling the universe can be regarded as ideal fluid. Among
alternative scenarios considered in the literature there are some
that relax that assumption, introducing a solid component of the
universe formed in the early stage of Friedmann expansion
\cite{{bs},{bbs},{bm1},{bm2},{bm},{bp},{kum}}. The solid is
supposed to have negative pressure to energy density ratio $w$; in
particular, it can consist of cosmic strings ($w = - 1/3$) or
domain walls ($w = - 2/3$). Such matter starts to influence the
dynamics of the universe at late times only and has no effect on
the evolution of perturbations during the hot universe period.

To obtain large-scale perturbations whose magnitude at
recombination differs from their magnitude at the end of
inflation, we need a solid with $w \ge 1/3$. A scenario with {\it
radiation-like solid} ($w = 1/3$) was considered in \cite{bsk},
where it was shown that the solid produces an additional term in
the gravitational potentials that can be large at the beginning
but decays afterwards. If one introduces {\it stiff solid} ($w
> 1/3$) instead, the character of the expansion of the universe
changes for a limited period and a question arises whether this
cannot cause a shift in the nondecaying part of the potentials, in
analogy to what we observe in a universe filled with ideal fluid
as it passes from one expansion regime to another due to a jump in
$w$. If so, the incorporation of the solid into the theory, with
the value of its shear modulus left free, would enlarge the
interval of admissible values of the {\it primordial} potential,
extending in such a way the parameter space of inflationary
scenarios.

A possible realization of stiff solid would be a system of equally
charged particles with anisotropic short-range interaction. By
using Yukawa potential, one obtains stiff {\it fluid}
\cite{{zel},{wal}}; however, if the potential is squeezed in some
direction and the particles are arranged into a lattice, the
system acquires nonzero transversal as well as longitudinal sheer
modulus with respect to that direction.

In order that a solid, radiation-like or stiff, has an effect on
large-scale perturbations, the solidification has to be {\it
anisotropic}, producing a solid with flat internal geometry and
nonzero shear stress. Such solidification might possibly take
place in case the Friedmann expansion was preceded by {\it solid
inflation}, driven by a solid with $w < 0$ rather than by a scalar
field \cite{{gru},{end},{bar},{sit}}.

In the paper we study how a stiff solid formed during Friedmann
expansion would influence the evolution of large-scale
perturbations. In section \ref{sec:pert} we derive solution for
such perturbations in a one-component universe and establish
matching conditions in a universe whose matter content has changed
abruptly; in section \ref{sec:stiff} we determine the behavior of
perturbations after the solid has been formed and find both
nondecaying and decaying part of Newtonian potential after
radiation prevailed again; and in section \ref{sec:con} we discuss
the results. Signature of the metric tensor is $(+ - - - )$ and a
system of units is used in which $c = 16\pi G = 1$.

%KAPITOLA 2

\section{Perturbations in the presence of solid}
 \label{sec:pert}

\subsection{Evolution equations}

Consider a flat FRWL universe filled with an elastic medium, fluid
or solid, with energy density $\rho$ and pressure $p$, and denote
the conformal time by $\eta$ and the scale parameter by $a$.
Expansion of the universe is described by the equations
\begin{equation}
a' = \Bl\frac 16 \r a^4\Br^{1/2}, \quad \r' = - 3\H \r_+,
 \label{eq:Eeq}
\end{equation}
where the prime denotes differentiation with respect to $\eta$,
$\H = a'/a$ and $\r_+ = \r + p$.

In a perturbed universe, spacetime metric and stress-energy tensor
acquire small space-depen\-dent corrections $\d g_{\mu \nu}$ and
$\d T_{\mu \nu}$. We will use the {\it proper-time gauge} in which
$\d g_{00} = 0$ (the cosmological time $t = \displaystyle \int a
d\eta$ coincides with the proper time of local observers). The
metric in this gauge is
\begin{equation}
ds^2\ \hat =\ a^2 [d\eta^2 + 2 B_{,i} d\eta dx^i - (\d_{ij} -
2\psi \d_{ij} - 2E_{,ij}) dx^i dx^j],
 \label{eq:ds2}
\end{equation}
where the effective equality indicates that only the scalar part
of the quantity in question is given. Suppose the matter filling
the universe has Euclidean internal geometry and contains no
entropy perturbations. The perturbation to $T_{\mu \nu}$ is then
given solely by the perturbation to $g_{\mu \nu}$ and the shift
vector of matter $\bm \xi$. We will use the remaining gauge
freedom to impose the condition $\bm {\xi} = 0$, so that our gauge
will be also {\it comoving}. In this gauge, the perturbation of
mass density $\d \r = \d {T_0}^0$, the energy flux density $S^i =
- {T_i}^0$ and the perturbation of stress tensor $\d \tau^{ij} =
\d {T_i}^j$ are \cite{bs}
\begin{equation}
\d \r = \r_+ (3\psi + \E), \quad S^i\ \hat =\ - \r_+ B_{,i}, \quad
\d \tau^{ij}\ \hat =\ - K (3\psi + \E) \d_{ij} - 2\mu E_{,ij}^T.
 \label{eq:dT}
\end{equation}
where $K$ is the {\it compressional modulus}, $\mu$ is the {\it
shear modulus} and the index `T' denotes the traceless part of the
matrix. (Our $K$ is 2 times greater and our $\mu$ is 4 times
greater than $K$ and $\mu$ in \cite{bs}. We have defined them so
in order to be consistent with the standard definitions in
Newtonian elasticity.)

The proper-time gauge is not defined uniquely since one can shift
the cosmological time by an arbitrary function $\d t({\bf x})$.
Under such shift, $E$ stays unaltered and $B$ and $\psi$ transform
as
\begin{equation*}
B \to B + \d \eta, \quad \psi \to \psi - \H \d \eta,
\end{equation*}
where $\d \eta = a^{-1} \d t$. This suggests that we represent $B$
and $\psi$ as
\begin{equation}
B = \B + \chi, \quad \psi = - \H \chi,
 \label{eq:Bps}
\end{equation}
where $\B$ stays unaltered by the time shift and $\chi$ transforms
as $\chi \to \chi + \d \eta$.

We will restrict ourselves to perturbations of the form of plane
waves with the wave vector $\bf k$, $\B$ and $\E \propto e^{i{\bf
k} \cdot {\bf x}}$. The action of the Laplacian then reduces to
the multiplication by $- k^2$; in particular, the definition of
$\E$ becomes $\E = - k^2 E$. For simplicity, we will suppress the
factor $e^{i{\bf k} \cdot {\bf x}}$ in $\B$ and $\E$, as well as
in other functions describing the perturbation. They will be
regarded as functions of $\eta$ only.

Evolution of scalar perturbations is governed by two differential
equations of first order for the functions $\B$ and $\E$, coming
from equations ${{T_i}^\mu}_{;\mu} = 0$ and $2G_{00} = T_{00}$.
The equations are \cite{pol}
\begin{equation}
\B' = (3c_{S0}^2 + \a - 1)\H \B + c_{S\|}^2 \E, \quad \E' = - (k^2
+ 3\a \H^2) \B - \a \H \E,
 \label{eq:dBdE}
\end{equation}
where $\a = \r_+/(2\H)^2 = (3/2) \r_+/\r$, $c_{S0}$ is the
``fluid'' sound speed (sound speed of the solid with suppressed
contribution of shear modulus), $c_{S0}^2 = K/\r_+$, and $c_{S\|}$
is the longitudinal sound speed, $c_{S\|}^2 = c_{S0}^2 + (4/3)
\mu/\r_+$. The only place where the shear modulus enters equations
(\ref{eq:dBdE}) is the term $c_{S\|}^2 \E$ in the equation for
$\B$.

Consider a one-component universe filled with a solid that has
both $p$ and $\mu$ proportional to $\r$. The quantity $K$ is then
proportional to $\r$, too, since $K = \r_+ c_{S0}^2$ and $c_{S0}^2
= dp/d\r$. Mechanical properties of such solid are given
completely by two dimensionless constants $w = p/\r$ and $\tm =
\mu/\r$. To simplify formulas, we will often use the constant $\b
= \mu/\r_+ = \tm/w_+$, where $w_+ = 1 + w$, instead of $\tm$.

For constant $w$ and $\tm$, the quantities appearing in the
equations for $\B$ and $\E$ are all constant, except for the
Hubble parameter that is proportional to $\eta^{-1}$. Explicitly,
\begin{equation*}
\a = \frac 32 w_+, \quad c_{S0}^2 = w, \quad c_{S\|}^2 = w + \frac
43 \b \equiv \tw, \quad \H = 2u \eta^{-1},
\end{equation*}
where $u = 1/(1 + 3w)$. With these expressions, equations for $\B$
and $\E$ simplify to
\begin{equation}
\B' = u (1 + 9w) \eta^{-1} \B + \tw \E, \quad \E' = - (k^2 + 18u^2
w_+ \eta^{-2}) \B - 3u w_+ \eta^{-1} \E,
 \label{eq:dBdEw}
\end{equation}
and after excluding $\E$, we arrive at an equation of second order
for $\B$,
\begin{equation}
\B'' + 2v \eta^{-1} \B' + [q^2 - (2v - b)\eta^{-2}] \B = 0,
 \label{eq:ddB}
\end{equation}
where $q = \sqrt{\tw}k$, $v = u(1 - 3w)$ and $b = 24 u^2 \tm$. The
equation is solved by Bessel functions of the argument $q\eta$,
multiplied by a certain power of $\eta$. We are interested only in
{\it large-scale perturbations}, that is, perturbations stretched
far beyond the sound horizon. Such perturbations have $q\eta \ll
1$, hence we can skip the term $q^2$ in the square brackets in
(\ref{eq:ddB}) to obtain
\begin{equation}
\B \doteq \eta (c_J \eta^{-m} + c_Y \eta^{-M}),
 \label{eq:asB}
\end{equation}
where the parameters $m$ and $M$ are defined in terms of the
parameters $\nu = v + 1/2 = (3/2) u (1 - w)$ and $n = \sqrt{\nu^2
- b}$ as $m = \nu - n$ and $M = \nu + n$. The constants are
denoted $c_J$ and $c_Y$ to remind us that the two terms in
(\ref{eq:asB}) come from the Bessel functions $J$ and $Y$.

The function $\B$ is non-oscillating for $b < \nu^2$ and
oscillating for $b > \nu^2$. Solutions of the second kind are well
defined if the solid was not present in the universe from the
beginning, but was formed at a finite time. Here we will restrict
ourselves to the solutions of the first kind, which means that we
will consider only values of the dimensionless shear stress $\tm
\le (3/32) (1 - w)^2$.

An approximate expression for $\E$ is obtained by inserting the
approximate expression for $\B$ into the first equation in
(\ref{eq:dBdEw}). In this way we find
\begin{equation}
\E \doteq \hc_J \eta^{-m} + \hc_Y \eta^{- M},
  \label{eq:asE}
\end{equation}
where $\hc_J$ and $\hc_Y$ are defined in terms of $c_J$ and $c_Y$
as $\hc_J = - (1/\tw) (3/2 - n) c_J$ and $\hc_Y = - (1/\tw) (3/2 +
n) c_Y$.

\subsection{Potentials $\Phi$ and $\Psi$}

Scalar perturbations we are interested in are most easily
interpreted in the {\it Newtonian gauge}, in which the metric is
\begin{equation}
ds^2\ \hat =\ a^2 [(1 + 2\Phi)d\eta^2  - (1 - 2\Psi) d{\bf x}^2].
 \label{eq:ds2N}
\end{equation}
Let us express the potentials $\Phi$ and $\Psi$ in terms of the
functions $\B$ and $\E$. If we perform explicitly the coordinate
transformation from the proper-time to Newtonian gauge, we find
(see equation (7.19) in \cite{mukh})
\begin{equation}
\Psi = \H (\B - E').
 \label{eq:Ps}
\end{equation}
For $\Phi$ we could proceed analogically, but it is simpler to use
Einstein equations. If we write the scalar part of the stress
tensor as a sum of pure trace and traceless part, $\tau^{ij}\ \hat
=\ \tau^{(1)} \d_{ij} + {\tau^{(2)T}}_{\mbox{\hskip -3.5mm},ij}$,
from equations $2G_{ij} = T_{ij}$ we obtain that the difference of
$\Phi$ and $\Psi$ is given by the latter quantity (see equation
(7.40) in \cite{mukh}),
\begin{equation*}
\dP \equiv \Phi - \Psi = \frac 12\tau^{(2)} a^2.
\end{equation*}
By inserting here from the third equation (\ref{eq:dT}) we find
\begin{equation}
\dP = - \mu a^2 E.
 \label{eq:DPh}
\end{equation}
We can see that in a universe filled with an ideal fluid ($\mu =
0$) the potentials $\Phi$ and $\Psi$ coincide.

After inserting into the expression for $\Psi$ from the second
equation in (\ref{eq:dBdE}) and into the expression for $\dP$ from
the first equation in (\ref{eq:Eeq}), we arrive at
\begin{equation}
\Psi = - k^{-2}\a  \H^2 (3 \H \B + \E), \quad \dP = 6\tm k^{-2}
\H^2 \E.
 \label{eq:PP}
\end{equation}
For the one-component universe introduced before, expressions for
$\Psi$ and $\dP$ become
\begin{equation}
\Psi = - 6u^2 w_+ (k\eta)^{-2}(6u \eta^{-1} \B + \E), \quad \dP =
24 u^2 \tm (k\eta)^{-2} \E.
 \label{eq:PPw}
\end{equation}
With $\B$ and $\E$ given in (\ref{eq:asB}) and (\ref{eq:asE}),
both $\Phi$ and $\Psi$ are linear combinations of $\eta^{-2 - m}$
and $\eta^{-2 - M}$. For an ideal fluid $m = 0$ and $M = 2\nu$, so
that we expect the function $\Phi$ to be linear combination of
$\eta^{-2}$ and $\eta^{-2\nu_+}$, where $\nu_+ = 1 + \nu$. This
is, however, not true because the coefficient in front of
$\eta^{-2}$ turns out to be zero. Thus, if we want to establish
how $\Phi$ looks like for an ideal fluid, or how $\Phi$ and $\Psi$
look like for a solid with small $\tm$, we must add the
next-to-leading term to the $J$-part of both expressions
(\ref{eq:asB}) and (\ref{eq:asE}). The term is suppressed by the
factor $(q\eta)^2$, therefore the $J$-part of $\Phi$ for an ideal
fluid is constant and the $J$-part of $\Phi$ and $\Psi$ for a
solid with small $\tm$ acquires a term proportional to
$\eta^{-m}$. For a universe filled with an ideal fluid we have
\begin{equation}
\B \doteq \eta (c_J + c_Y \eta^{-2\nu}), \quad \E \doteq \hc_J +
\hc_Y \eta^{- 2\nu},
 \label{eq:asBEf}
\end{equation}
where $\hc_J$ and $\hc_Y$ are defined in terms of $c_J$ and $c_Y$
as $\hc_J = - 6u c_J$ and $\hc_Y = - 3u (w_+/w) c_Y$. After
computing the additional terms in $\B$ and $\E$ and inserting the
resulting expressions into equations (\ref{eq:PPw}), we arrive at
\begin{equation}
\Phi \doteq C_J + C_Y \eta^{-2\nu_+},
 \label{eq:asPf}
\end{equation}
where $C_J$ and $C_Y$ are defined in terms of $c_J$ and $c_Y$ as
$C_J = 3u^2 (w_+/\nu_+) c_J$ and $C_Y =$ \linebreak $12u^2 w_+ \nu
q^{-2} c_Y$.

\subsection{Transitions with jump in $w$ and $\tm$}

Suppose the functions $w_\eta$ and $\tm_\eta$ change at the given
moment $\eta_{tr}$ (``transition time'') from $(w_I, \tm_I)$ to
$(w_\I, \tm_\I) = (w_I + \D w, \tm_I + \D \tm)$. (We have attached
the index $\eta$ to the symbols $w$ and $\tm$ in order to
distinguish the functions denoted by them from the values these
functions assume in a particular era.) Rewrite the first equation
in (\ref{eq:dBdE}) as
\begin{equation}
\B' = c_{S0}^2 (3\H \B + \E) + \Bl\frac 32 w_{\eta +} - 1\Br\H \B
+ \frac 43 \b_\eta \E,
 \label{eq:Brew}
\end{equation}
where
\begin{equation}
c_{S0}^2 = \frac {dp}{d\r} = w_\eta + \r \frac {dw_\eta}{d\r}.
 \label{eq:cs02}
\end{equation}
Because of the jump in $w_\eta$, there appears $\d$-function in
$c_{S0}^2$, and to account for it we must assume that $\B$ has a
jump, too. However, on the right hand side of equation
(\ref{eq:Brew}) we then obtain an expression of the form
``$\theta$-function$\tim \d$-function''; and if we rewrite $\B'$
as
\begin{equation*}
\B' = \frac {d\B}{d\r} \r' = - 3\H \r w_{\eta +} \frac {d\B}{d\r},
\end{equation*}
on the left hand side there appears another such expression. To
give meaning to the equation we must suppose that $w_\eta$ changes
from $w_I$ to $w_\I$ within an interval of the length $\D \r \ll
\r_{tr}$, and send $\D \r$ to zero in the end. If we retain just
the leading terms in equation (\ref{eq:Brew}) in the interval with
variable $w$, we obtain
\begin{equation}
w_{\eta +} \frac {d\B}{d\r} = -\Bl \B + \frac
{\E_{tr}}{3\H_{tr}}\Br \frac {dw_\eta}{d\r},
 \label{eq:Blead}
\end{equation}
where we have used the fact that, as seen from the second equation
in (\ref{eq:dBdE}), the function $\E$ is continuous at $\eta =
\eta_{tr}$. The solution is
\begin{equation*}
\B + \frac {\E_{tr}}{3\H_{tr}} = \frac \C{w_{\eta +}}.
\end{equation*}
Denote the jump of the function at the moment $\eta_s$ by square
brackets. To determine $[\B]$, we express $\B_I$ and $\B_\I$ in
terms of $w_{I +}$ and $w_{\I +}$, compute the difference $\B_\I -
\B_I$ and use the expression for $\B_I$ to exclude $\C$. In this
way we find
\begin{equation}
[\B] = - \frac {\D w}{w_{\I +}} \Bl \B_I + \frac
{\E_{tr}}{3\H_{tr}} \Br.
 \label{eq:jB}
\end{equation}
Note that the same formula is obtained if we assume that the
functions with jump are equal to the mean of their limits from the
left and from the right at the point where the jump occurs.

To justify the expression for $[\B]$, let us compute the jump in
$\Psi$. It holds
\begin{equation*}
[\Psi] =  - \frac 32 k^{-2} \H_{tr}^2 (3 \H_{tr} [w_{\eta+} \B] +
\D w \E_{tr}),
\end{equation*}
and if we write $[w_{\eta+} \B] = w_{\I+} [\B] + \D w \B_I$ and
insert for $[\B]$, we find that $[\Psi]$ vanishes. This must be so
because for $\Psi$ we have (see equation (7.40) in \cite{mukh})
\begin{equation*}
\Psi'' +  \H (2\Psi' + \Phi') + (2\H' + \H^2) \Psi = - \frac
14\overline{\d \tau^{(1)}},
\end{equation*}
where the bar indicates that the quantity $\d \tau^{(1)}$ is
computed in Newtonian gauge. A jump in $\Psi$ would produce a
derivative of $\d$-function in the first term, but no such
expression with opposite sign appears in the other terms.

The jump in $\B'$ can be found from equation (\ref{eq:Brew}) by
computing the jump of the right hand side, with no need for the
limiting procedure we have used when determining the jump in $\B$.
The result is
\begin{equation}
[\B'] = 4 \frac {\D w}{w_{\I+}} \H_{tr} \B_{tr} + \Bl \frac {5 -
3w_\I}{6w_{\I+}} \D w + \frac 43 \D \b \Br \E_{tr}.
 \label{eq:jdB}
\end{equation}

\section{Scenario with stiff solid}
 \label{sec:stiff}

\subsection{Expansion of the universe}

Suppose at some moment $\eta_s$ the hot universe underwent a phase
transition during which a part of radiation ($w = 1/3$)
instantaneously turned into a stiff solid ($w > 1/3$). In a
one-component universe with given parameter $w$, the density of
matter falls down the faster the greater the value of $w$. As a
result, if the solid acquired a substantial part of the energy of
radiation at the moment it was formed, it dominated the evolution
of the universe for a limited period until radiation took over
again. Let us determine the function $a(\eta)$ for such universe.

Denote the part of the total energy that remained stored in
radiation after the moment $\eta_s$ by $\e$. In the period with
pure radiation ($\eta < \eta_s$) the mass density was $\r = \r_s
(a_s/a)^4$, so that from the first equation in (\ref{eq:Eeq}) we
obtain
\begin{equation}
a = C \eta, \quad C = \Bl\frac 16 \r_s a_s^4\Br^{1/2}.
 \label{eq:arad}
\end{equation}
In the period with a mix of radiation and solid ($\eta > \eta_s$)
the mass density is
\begin{equation*}
\r = \e \r_s (a_s/a)^4  + (1 - \e) \r_s (a_s/a)^{3w_+} = \r_s
(a_s/a)^4 [\e + (1 - \e)(a_s/a)^\D],
\end{equation*}
where $\D = 3w_+ - 4$. As a result, the first equation in
(\ref{eq:Eeq}) transforms into
\begin{equation}
a' = C [\e + (1 - \e)(a_s/a)^\D]^{1/2}.
 \label{eq:aseq}
\end{equation}
For $w > 1/3$ the parameter $\D$ is positive, therefore the second
term eventually becomes less than the first term even if $\e \ll
1$.

Suppose radiation retained less than one half of the total energy
at the moment of radiation-to-solid transition ($\e < 1/2$). The
subsequent expansion of the universe can be divided into two eras,
solid-dominated and radiation-dominated, separated by the time
$\eta_{rad}$ at which the mass densities of the solid and
radiation were the same. The value of $\eta_{rad}$ is given by
\begin{equation}
a_{rad} = a_s (\e^{-1} - 1)^{1/\D}.
 \label{eq:aeq}
\end{equation}

Suppose now that the post-transitional share of energy stored in
radiation was small ($\e \ll 1$). The universe then expands by a
large factor between the times $\eta_s$ and $\eta_{rad}$,
\begin{equation*}
a_{rad} \doteq a_s \e^{-1/\D} \gg a_s,
\end{equation*}
and can be described in a good approximation as if it was filled
first with pure solid and then with pure radiation. Thus, equation
(\ref{eq:aseq}) can be replaced by
\begin{equation}
a' \doteq \bigg\{ \mbox{\hskip -2mm}
  \left. \begin{array} {l}
  C (a_s/a)^{\D/2} \mbox{ for } \eta < \eta_{rad}\\
  \sqrt{\e}C \mbox{ for } \eta > \eta_{rad}\\
  \end{array}\mbox{\hskip -1mm}. \right.
 \label{eq:aseqap}
\end{equation}
The solution is
\begin{equation}
a \doteq \bigg\{ \mbox{\hskip -2mm}
  \left. \begin{array} {l}
  \big[(\D/2 + 1) a_s^{\D/2} C\te\big]^{\frac 1{\D/2 + 1}} \mbox{ for }
  \eta < \eta_{rad}\\
  \sqrt{\e} C \tte \mbox{ for } \eta > \eta_{rad}\\
  \end{array}\mbox{\hskip -1mm}, \right.
 \label{eq:asap}
\end{equation}
where $\te$ and $\tte$ are shifted time variables, $\te = \eta -
\eta_*$ and $\tte = \te - \eta_{**}$. From the approximate
expression for $a_{rad}$ we obtain
\begin{equation}
\te_{rad} = \frac 1{\D/2 + 1} \e^{- \frac {\D/2 + 1}\D} \eta_s,
 \label{eq:eeq}
\end{equation}
and by matching the solutions at $\eta_s$ and $\eta_{rad}$ we find
\begin{equation}
\eta_* = \frac {\D/2}{\D/2 + 1} \eta_s, \quad \eta_{**} = - \frac
\D2 \te_{rad},
 \label{eq:eestar}
\end{equation}

Note that equation (\ref{eq:aseq}) solves analytically for $w =
2/3$ and $w = 1$, when $\D = 1$ and $\D = 2$. We do not give these
solutions here since will not need them in what follows.

The two equations in (\ref{eq:eestar}) can be rewritten to
formulas for the ratios of shifted and unshifted times,
\begin{equation*}
\frac {\te_s}{\eta_s} = \frac 1{\D/2 + 1} = \frac u{u_0}, \quad
\frac {\tte_{rad}}{\te_{rad}} = \frac \D2 + 1 = \frac {u_0}u,
\end{equation*}
where $u_0$ is the value of $u$ in the radiation-dominated era,
$u_0 = 1/2$. These equations stay valid also after we replace
radiation by an ideal fluid with an arbitrary pressure to energy
density ratio $w_0$. To demonstrate that, let us derive them from
the condition of continuity of the Hubble parameter. If the
universe is filled in the given period with matter with the given
value of $w$, its scale parameter depends on a suitably shifted
time $\te$ as $a \propto \te^{2u}$. Thus, its Hubble parameter is
$\H = 2u \te^{-1}$ and the requirement that $\H$ is continuous at
the moment when $w$ changes from $w_I$ to $w_\I$ is equivalent to
$\te_\I/\te_I = u_\I/u_I$.

\subsection{Behavior of the function $\B$}

We are interested in large-scale perturbations in a universe in
which the parameters $w$ and $\tm$ assume values $(w_0, 0)$ before
$\eta_s$, $(w, \tm)$ between $\eta_s$ and $\eta_{rad}$, and $(w_0,
0)$ after $\eta_{rad}$. (Most of the time we will leave $w_0$
free, only at the end we will put $w_0 = 1/3$.) Denote the
functions describing the perturbation before $\eta_s$ and after
$\eta_{rad}$ by the indices 0 and 1 respectively, and keep the
functions referring to the interval between $\eta_s$ and
$\eta_{rad}$ without index. If only the nondecaying part of the
perturbation (the part with constant $\Phi$) survives at the
moment $\eta_s$, the functions $\B_0$ and $\E_0$ can be replaced
by their $J$-parts,
\begin{equation}
\B_0 = c_{J0} \eta, \quad \E_0 = \hc_{J0} = - 6u_0 c_{J0}.
 \label{eq:B0E0}
\end{equation}
For the functions $\B$ and $\E$ we have expressions (\ref{eq:asB})
and (\ref{eq:asE}) with $\eta$ replaced by $\te$ and for the
function $\B_1$ we have the first equation (\ref{eq:asBEf}) with
$c_J$ and $c_Y$ replaced by $c_{J1}$ and $c_{Y1}$, $\nu$ replaced
by $\nu_0$ and $\eta$ replaced by $\tte$. All we need to obtain
the complete description of the perturbation is to match the
expressions for $\B_0$, $\B$ and $\B_1$ with the help of the
expressions for $\E_0$ and $\E$ at the moments $\eta_s$ and
$\eta_{rad}$.

At the moment $\eta_s$, the jumps in $w_\eta$ and $\tm_\eta$ are
$\D w_s = w - w_0 \equiv \D w$ and $\D \tm_s = \tm$. By using
these values and the identity $\E_0 = -3\H_s \B_{0s}$, we find
\begin{equation*}
[\B]_s = 0, \quad [\B']_s = - \Bl \frac 12 \D w - \frac 43 \b \Br
\E_0,
\end{equation*}
Denote $x_0 = c_{J0}$. Equations for the unknowns $\tx = c_J
\te_s^{-m}$ and $\ty = c_Y \te_s^{-M}$ are
\begin{equation}
\tx + \ty = \frac {u_0}u x_0, \quad (1 - m) \tx + (1 - M) \ty =
\BL 1 + 8u_0 \Bl \frac 38 \D w  - \b \Br\BR x_0,
 \label{eq:tmatch}
\end{equation}
and their solution is
\begin{equation}
\tx = \frac {u_0}u \frac 1{2n} (M - 8u \b) x_0, \quad \ty = -
\frac {u_0}u \frac 1{2n} (m - 8u \b) x_0.
 \label{eq:tsolxy}
\end{equation}

At the moment $\eta_{rad}$, the jumps in $w_\eta$ and $\tm_\eta$
are $\D w_{rad} = - \D w$ and $\D \b_{rad} = - \tm$. By inserting
these values into the expressions for $[\B]$ and $[\B']$ we obtain
\begin{equation*}
[\B]_{rad} =  \frac {\D w}{w_{0+}} \Bl \B_{rad} + \frac
{\E_{rad}}{3\H_{rad}} \Br, \quad [\B']_{rad} = -4 \frac {\D
w}{w_{0+}} \H_{rad} \B_{rad} - \Bl \frac {5 - 3w_0}{6w_{0+}} \D w
+ \frac 43 \b \Br \E_{rad}.
\end{equation*}
Introduce the constants
\begin{equation}
\tX = c_J \te_{rad}^{-m} = p^{-m} \tx, \quad \tY = c_Y
\te_{rad}^{-M} = p^{-M} \ty,
 \label{eq:tXtY}
\end{equation}
where $p$ is the ratio of final and initial moments of the period
during which the solid affects the dynamics of the universe, $p =
\te_{rad}/\te_s$. Equations for the unknowns $\ttx = c_{J1}$ and
$\tty = c_{Y1} \tte_{rad}^{-2\nu_0}$ are
\begin{equation}
\ttx + \tty = \frac u{u_0} (K_J\tX + K_Y\tY), \quad \ttx + (1 -
2\nu_0) \tty = L_J\tX + L_Y\tY,
 \label{eq:ttmatch}
\end{equation}
where the coefficients on the right hand side are defined as
\begin{equation*}
K_J = \frac 1{w_{0+}} \BL w_+ - \frac {\D w}{6u\tw} (m + 6uw) \BR,
\quad K_Y = \mbox{ditto with } m \to M,
\end{equation*}
and
\begin{equation*}
L_J = 1 - m - \frac {8u\D w}{w_{0+}} + \frac {m + 6uw}{\tw} \Bl
\frac{5 - 3w_0}{6w_{0+}} \D w + \frac 43 \tm \Br, \quad L_Y =
\mbox{ditto with } m \to M,
\end{equation*}
The solution is
\begin{equation}
\ttx = \frac 1{2\nu_0} (M_J \tX + M_Y \tY). \quad \tty = - \frac
1{2\nu_0} (N_J \tX + N_Y \tY)
 \label{eq:ttxy}
\end{equation}
with the constants $M_\a$ and $N_\a$, $\a = J$, $Y$, defined in
terms of the constants $L_\a$ and $K_\a$ as
\begin{equation*}
M_\a = L_\a - (1 - 2\nu_0) \frac u{u_0} K_\a, \quad N_\a = L_\a -
\frac u{u_0} K_\a.
\end{equation*}

\subsection{Behavior of potentials}

Knowing how the function $\B$ looks like, we can establish the
time dependence of the Newtonian potential $\Phi$ and the
potential describing the curvature of 3-space $\Psi$. Before the
time $\eta_s$, both potentials are the same, $\Phi_0$ as well as
$\Psi_0 = C_{J0} \sim x_0$. Between the times $\eta_s$ and
$\eta_{rad}$, the potentials are given by the two equations in
(\ref{eq:PPw}) with $\eta$ replaced by $\te$. With $\B$ and $\E$
inserted from equations (\ref{eq:asB}) and (\ref{eq:asE}), both
$\Phi$ and $\Psi$ become sums of terms proportional to $\te^{-2 -
m}$ and $\te^{-2 - M}$. We have already mentioned that for $\tm =
0$ the coefficient in the first term in $\Phi = \Psi$ is zero, and
one easily verifies that for $w > 1/3$ and $\tm$ close to zero the
first coefficient in both $\Phi$ and $\Psi$ is proportional to
$\tm$. (After a simple algebra we find that it is proportional to
$m(1 - 4\b) - 8\b$ and $m - 8u\b$ for $\Phi$ and $\Psi$
respectively, with $m$ reducing to $b/(2\nu) = 8uw_+\b/(1 - w)$ in
the limit $\b \ll 1$.) The coefficients contain the constants
$c_J$ and $c_Y$ and if we use $c_J \propto \tx$ and $c_Y \propto
\ty$ with $\tx$ and $\ty$ given in equation (\ref{eq:tsolxy}), we
find that the second coefficient is proportional to $\tm$, too.
(In the expression for $\ty$ we encounter the factor $m - 8u\b$
again.) Both coefficients contain also the factor $x_0 \sim \Pz$,
therefore for $\eta$ close to $\eta_s$ we have $\Phi$ as well as
$\Psi \sim \tm (k \te)^{-2} \Pz$. As $\eta$ grows, the first
correction to the term proportional to $\te^{-2 - m}$, which is of
order $\Pz$, may take over while the perturbation still remains
stretched over the horizon. However, in order that our
approximation is valid, this term must be negligible in the first
period after the moment $\eta_s$. (Note that this does not hold
for the potential $\Psi$ just after $\eta_s$: it equals $\Pz$ at
$\eta_s$, hence it is dominated by the correction term for a short
period afterwards.) As a result, $\tm$ must be not {\it too} close
to zero, $\tm \gg (k \te_s)^2$.

For large enough $\tm$, $\Phi$ and $\Psi$ can become much greater
in absolute value not only than $\Pz$, but also than 1. The theory
then seems to collapse, but it does not because, as can be checked
by direct computation, $kB$, $\psi$ and $\E$ remain much less than
1. (A detailed discussion for $\D w = 0$ can be found in
\cite{bsk0}.) Thus, the proper-time comoving gauge which we have
implemented instead of more common, and intuitively more
appealing, Newtonian gauge, is not only convenient
computationally, but also preferable on principal grounds. Without
it we would not know that the perturbations stay small and the
linearized theory stays applicable after a solid with
above-critical parameter $\tm$ was formed, causing the potentials
$\Phi$ and $\Psi$ to rise beyond control.

We are interested in the potential $\Phi$ after the moment
$\eta_{rad}$, when both potentials coincide again. Denote the
nondecaying part of $\Phi$ in that period as $\Po$. It holds $\Po
= C_{J1}$, and by using the relation between $C_J$ and $c_J$ we
obtain
\begin{equation}
\Po = 3u_0^2 \frac {w_{0+}}{\nu_{0+}} \ttx.
 \label{eq:P10}
\end{equation}
Here we must insert for $\ttx$ from equation (\ref{eq:ttxy}), with
$\tX$ and $\tY$ given in equation (\ref{eq:tXtY}), $\tx$ and $\ty$
given in equation (\ref{eq:tsolxy}) and $x_0$ given by
\begin{equation*}
\Pz = 3u_0^2 \frac {w_{0+}}{\nu_{0+}} x_0.
\end{equation*}
The resulting expression for $\Po$ is
\begin{equation}
\Po = \frac 1{2\nu_0} \frac {u_0}u \frac 1{2n} (\hM_J p^{-m} -
\hM_Y p^{-M}) \Pz,
 \label{eq:P1}
\end{equation}
with the coefficients $\hM_J$ and $\hM_Y$ defined as
\begin{equation*}
\hM_J = M_J (M - 8u \b), \quad \hM_Y = M_Y (m - 8u \b).
\end{equation*}
After some algebra the coefficients reduce to
\begin{equation}
\hM_J = 2\nu_0 \frac u{u_0} M - b, \quad \hM_Y = \mbox{ditto with
} M \to m.
 \label{eq:MJMY}
\end{equation}

Let us now determine how fast the function $\Phi$ approaches its
limit value. Denote $\ttz = q_0 \tte$, where $q_0 = \sqrt{w_0} k$.
The decaying part of $\Phi$ in the period under consideration is
\begin{equation}
\D \Phi_1 = -2\nu_{0+} \frac {u_0}u \frac 1{2n} (\hN_J p^{-m} -
\hN_Y p^{-M}) \ttz_{rad}^{-2} \z^{-2 \nu_{0+}} \Pz,
 \label{eq:DP1}
\end{equation}
where $\z$ is rescaled time normalized to 1 at the moment
$\eta_{rad}$, $\z = \tte/\tte_{rad}$, and the coefficients $\hN_J$
and $\hN_Y$ are defined in terms of $N_J$ and $N_Y$ in the same
way as the coefficients $\hM_J$ and $\hM_Y$ in terms of $M_J$ and
$M_Y$. After rewriting the former coefficients similarly as we did
with the latter ones, we obtain
\begin{equation}
\hN_J = \hN_Y = -\frac {w_0}{w_{0+}} 2b.
 \label{eq:NJNY}
\end{equation}
From these equations and equations (\ref{eq:P1}) and
(\ref{eq:MJMY}) we find that the ratio of the decaying and
nondecaying part of $\Phi$ at the moment of solid-to-radiation
transition is
\begin{equation}
\left. \frac {\Delta \Phi_1}{\Po} \right|_{rad} = R
\ttz_{rad}^{-2}, \quad R = 4\nu_{0} \nu_{0+} \frac {w_0}{w_{0+}}
\frac{2u_0 b}{2\nu_0 u [n \coth (n \log p) + \nu] - u_0 b}.
 \label{eq:Rrad}
\end{equation}
The ratio is greater than one for $\b \gtrsim \ttz_{rad}^2$. The
function $\Phi$ is then dominated by the decaying term at the
moment $\eta_{rad}$, the nondecaying term taking over later, at
the moment $\eta_{nd}$ given by
\begin{equation}
\ttz_{nd} = R^{\frac 1{2\nu_{0+}}} \ttz_{rad}^{1 - \frac
1{\nu_{0+}}}.
 \label{eq:zdec}
\end{equation}
The exponent at $\ttz_{rad}$ is positive for any $w_0 < 1$ (it
equals 1/3 for $w_0 = 1/3$) and the constant $R$ is of order 1 or
less. Thus, if the perturbation was stretched over the horizon at
the moment the fluid originally filling the universe started to be
dominating again ($\ttz_{rad} \ll 1$), it will stay so at the
moment the nondecaying term prevails over the decaying one
($\ttz_{nd} \ll 1$).

The time $\eta_{rad}$ must not be too close to the time of
recombination $\eta_{re}$, if the spectrum of large-angle CMB
anisotropies is not to be tilted too much. If we denote the wave
number of perturbations crossing the sound horizon at
recombination as $k^\0$, the perturbations with the longest
wavelength that can be observed in CMB have $k \sim 0.01 k^\0$.
For $w_0 = 1/3$, Newtonian potential after the moment $\eta_{rad}$
is $\Phi_1 = (1 + R \ttz_{rad}^{-2} \z^3) \Po = (1 + R \ttz_{rad}
\ttz^{-3}) \Po$, and if we take into account that the value of
$\ttz^\0_{re}$ is approximately 1, we find
\begin{equation*}
r \equiv \left. \frac {\Phi_1 (k^\0)}{\Phi_1 (0.01 k^\0)}
\right|_{re} = \frac {1 + R \ttz_{rad}^\0}{1 + 10^4 R
\ttz_{rad}^\0} \doteq 1 - 10^4 R \ttz_{rad}^\0.
\end{equation*}
The observational value of $r$ is 0.01$^{n_S - 1}$, where $n_S$ is
the {\it scalar spectral index}, a characteristic of perturbations
whose deviation from 1 (about $-0,04$ according to observations)
describes the tilt of the scalar spectrum. If we allow for a tilt
of the primordial spectrum, too, the expression for $r$ must be
multiplied by 0.01$^{n_{S0} - 1}$. Denote $p_* = 1/\ttz_{rad}^\0 =
\tte_{re}/\tte_{rad} = a_{re}/a_{rad} = T_{rad}/T_{re}$ and
require that $n_S$ differs from $n_{S0}$ at most by some $\D n_S
\ll 1$. To ensure that, $p_*$ must satisfy
\begin{equation}
p_* > 2 \times 10^3 R \D n_S^{-1}.
\end{equation}

For numerical calculations we need the value of $p$. It is a ratio
of {\it times}, but can be rewritten in terms of a ratio of {\it
scale parameters} or {\it temperatures}, $P = a_{rad}/a_s =
T_s/T_{rad}$, as
\begin{equation}
p = P^{\frac 1{2u}}.
 \label{eq:pP}
\end{equation}
The value of $p$, or equivalently, $P$, determines the interval of
admissible $w$'s. To obtain it, note that for $w_0 = 1/3$ equation
(\ref{eq:aeq}) yields $P = (\e^{-1} - 1)^{1/\D} \doteq
\e^{-1/\D}$, or
\begin{equation}
P \doteq \e^{- \frac 1{3\D w}}.
 \label{eq:Papp}
\end{equation}
(This is consistent with equation (\ref{eq:eeq}), which can be
rewritten as $p = \e^{- \frac {\D/2 + 1}\D} = \e^{- \frac 1{6u\D
w}}$.) Thus, the jump in the parameter $w$ for the given ratio $P$
must satisfy
\begin{equation}
\D w \doteq \frac {\log 1/\e}{3 \log P} \gtrsim \frac 1{3\log P}.
 \label{eq:Dw}
\end{equation}

The dependence of the quantities $\phi = \Po/\Pz$ and $R$ on the
parameter $\b$ is depicted in fig.~1.
\begin{figure}[ht]
\centerline{\includegraphics[width=0.8\textwidth]{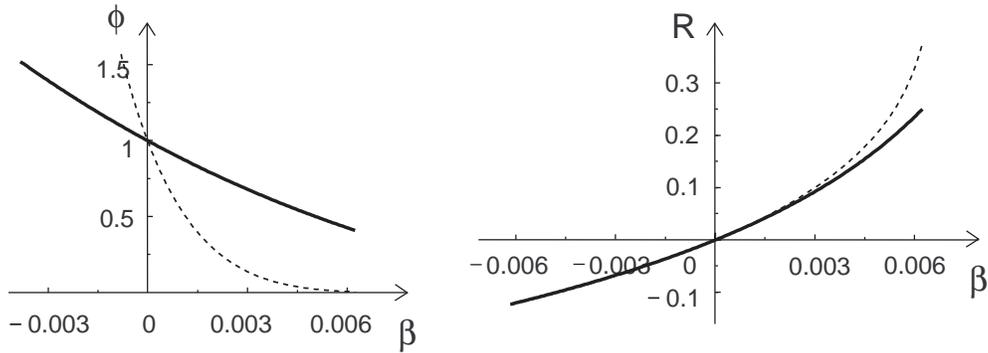}}
\centerline{\parbox {13.5cm}{\caption{\small Final value of
Newtonian potential in a universe with stiff solid (left) and
normalized ratio of decaying to nondecaying part of the potential
at solid-to-radiation transition (right), plotted as functions of
dimensionless shear modulus}}}
  \label{fig:PP0}
\end{figure}
The values of $w_0$ and $w$ are 1/3 and 2/3 on both panels and the
solid and dotted lines correspond to $P = 10^3$ and $P = 10^{13}$
respectively. The lines are terminated at $\b = 1/160$, which is
the maximum value of $\b$ admitting non-oscillating solutions in a
solid with $w = 2/3$.

For completeness, we have included also values $\b < 0$ into the
graphs. The transversal sound speed squared is negative for such
$\b$, so that the vector perturbations start to grow exponentially
once they have appeared. As a result, the theory is acceptable
only if such perturbations are produced neither during inflation
(which is the case in simplest models) nor in the subsequent phase
transitions.

The parameter $P$ assumes the smaller value if, for example, the
solid dominated the dynamics of the universe between the
electroweak and confinement scale, and the greater value, if the
solid was formed as soon as at the GUT scale and dominated the
dynamics of the universe up to the electroweak scale. Unless the
parameter $w$ of the solid is close to that of radiation, the
fraction of energy which remains stored in radiation after the
solid has been formed must be quite small in the former case and
very small in the latter case. For $w = 2/3$ this fraction equals
$1/P$, so that for the greater $P$ the mechanism of the
radiation-to-solid transition must transfer to the solid all but
one part in 10 trillions of the energy of radiation.

The quantity $\phi$ is the factor by which the value of the
potential $\Phi$ changes due to the presence of stiff solid in the
early universe. From the left panel of fig. 1 we can see that
$\Phi$ is shifted upwards for $\b < 0$ and downwards for $\b > 0$,
and the enhancement factor decreases monotonically with $\b$, the
steeper the larger the value of $P$. For maximum $\b$ the function
$\Phi$ is suppressed by the factor 0.41 if $P = 10^3$ and by the
factor 0.004 if $P = 10^{13}$. The quantity $R$ determines,
together with the parameter $\D n_S$, the minimal duration of the
period between the moment when radiation took over again and
recombination. According to the right panel of fig. 1, the
temperature at the beginning of this period had to be at least $8
\times 10^3 \D n_S^{-1} T_{re} \doteq 0.2\ (\D n_S/0.01)^{-1}$ MeV
for maximum $\b$ and $P = 10^3$.

%KAPITOLA 4

\section{Conclusion}
 \label{sec:con}

We have studied a scenario with stiff solid appearing in the hot
universe and dominating the evolution of the universe during a
limited period before recombination. In comparison with the
scenario containing radiation-like solid \cite{bsk}, a new effect
is that the nondecaying part of Newtonian potential becomes
suppressed. This might raise hope that the tensor-to-scalar ratio
is enhanced, which would surely be interesting from the
observational point of view. However, a straightforward
calculation shows that the tensor perturbations are suppressed by
exactly the same factor as the scalar ones. The shift in Newtonian
potential towards less values means that the rms of primordial
potential was in fact greater than supposed. As a result, there
appears an additional freedom in the choice of the parameters of
inflaton potential; for example, one can use potentials with
smaller inclination of the plateau than in the case without solid
when implementing slow-roll inflation.

\end{document}